\documentclass[prl,twocolumn, amsmath,amssymb,
reprint,%
author-year,%
author-numerical,%
dvipdfmx
]{revtex4}

\usepackage{graphicx}
\usepackage{bm}
\usepackage{color}

\begin{document}


\title{High-throughput computational search for two-dimensional binary compounds: Energetic stability versus synthesizability of three-dimensional counterparts}

\author{Shota Ono}
\email{shota\_o@gifu-u.ac.jp}
\author{Honoka Satomi}
\affiliation{Department of Electrical, Electronic and Computer Engineering, Gifu University, Gifu 501-1193, Japan}

\begin{abstract}
Using first principles calculations, the energetic stability of two-dimensional (2D) binary compounds $XY$ is investigated, where $X$ and $Y$ indicate the metallic element from Li to Pb in the periodic table. Here, 1081 compounds in the buckled honeycomb (BHC), buckled square, B2, L1$_0$, and B$_h$ structures are studied. For the compounds that have negative formation energy in the BHC structure or the compounds that can have the B$_h$ structure, the phonon dispersions of the 2D structures are also calculated. We demonstrate that (i) a negative formation energy is neither a sufficient nor necessary condition for yielding the dynamical stability of 2D compounds; and (ii) if a compound in the B$_h$ structure has been synthesized experimentally, that in the BHC structure is dynamically stable. 
\end{abstract}

\maketitle

{\it Introduction.} Recent advances have expanded the family of two-dimensional (2D) materials including graphene, black phosphorene, and transition metal dichalchogenides, and so forth, which can have a strong impact on scientific and technological innovation due to their electronic, mechanical, and optical properties \cite{ajayan}. Despite its growing members in the 2D materials, design principles for creating 2D materials have not been established. Intuitively, the structure of stable 2D materials is a counterpart of stable three-dimensional (3D) materials. For example, silicene is an atomically thin layer of silicon atoms constructed from the (111) surface of a silicon diamond structure \cite{cahangirov}. The validity of this concept has been confirmed in 2D metals including gallenene (Ga) \cite{kochat}, bismuthene (Bi) \cite{akturk}, poloniumene (Po) \cite{ono2020_Po}, and most elements in the periodic table \cite{nevalaita,hwang,ono2020}: For noble metals, the monolayer hexagonal and buckled honeycomb (BHC) structures are dynamically stable, as they can be constructed from the surfaces of fcc(111) and/or hcp(0001). However, it remains unclear whether this rule can be applied to compounds (ordered structures that consist of more than two elements). Recently, it has been reported that CuAu can have a 2D structure, where hexagonal layers of Cu and Au are stacked to form a bilayer structure \cite{zagler}, while CuAu has been known to have an L1$_0$ structure. 

As the computational capacity is increasing with time, high-throughput (HT) density-functional theory (DFT) methods enable us to carry out materials design \cite{schlender}. For example, it has been applied to identify new phases of binary compounds based on platinum-group metals \cite{hart} and high-entropy alloys \cite{troparevsky}. Recent HT-DFT studies have proposed various criteria that identify the stability of 2D materials. For example, the exfoliation energy of 2D layers \cite{ashton} and the relative difference between experimental lattice constants and DFT based lattice constants \cite{choudhary} are used to predict possible 2D materials.   

\begin{figure}[tt]
\center
\includegraphics[scale=0.42]{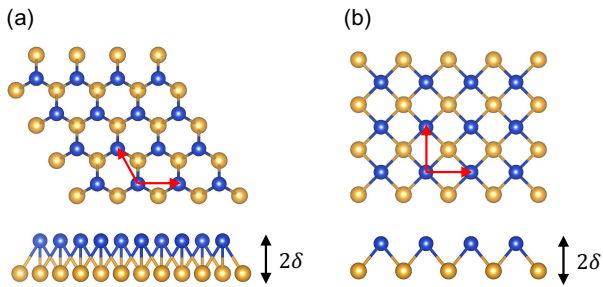}
\caption{Schematic illustration for binary compounds in the (a) BHC and (b) BSQ structures. The buckling height is expressed by $\delta$ (total thickness $2\delta$). The primitive lattice vectors are indicated by arrows (red). } \label{fig1} 
\end{figure}

In this Letter, by using HT-DFT, we search for 2D binary compounds that are energetically and dynamically stable. We investigate the energetic stability of 1081 binary compounds, created from a combination of 46 metallic elements (see Supplemental Material (SM)\cite{SM}), in the BHC and buckled square (BSQ) structures (see Fig.~\ref{fig1}). Among the many 2D structures \cite{balendhran,wang}, we have chosen BHC and BSQ because these structures for compounds are natural extensions of those for simple metals \cite{hwang,ono2020}, and the former BHC has been realized in 2D CuAu \cite{zagler}. For comparison, B2, L1$_0$, and B$_h$ structures are also studied, which correspond to bcc, fcc, and hcp structures in simple metals, respectively. Below, we demonstrate that (i) a negative formation energy is neither a sufficient nor necessary condition for producing the dynamical stability of 2D compounds; and (ii) if a compound in the B$_h$ structure is synthesizable, that in the BHC structure is dynamically stable. In contrast, if a compound in the BHC structure is unstable, that in the B$_h$ structure has not yet been synthesized. 


One of the open questions in materials science is to predict which structures can be created experimentally, i.e., {\it the synthesizability}. In general, the presence of metastable structures created experimentally can be rationalized by not only the energetic stability but also the overall structure of the potential energy surface, characterized by many degrees of freedom (i.e., the atom positions) \cite{ewels,sun,de}. For example, high energy barriers are necessary for preventing the metastable structure from transforming into more stable structures against perturbations \cite{ewels}. As an alternative approach for the energetic stability analysis, we regard 2D structures as building blocks for constructing 3D structures, e.g., the hexagonal lattice as a building block for BHC and B$_h$. We expect that such an interrelationship between metastable structures allow us to understand the synthesizability in detail.


{\it Computational details.} We calculate the total energy of binary compounds based on DFT implemented in \texttt{Quantum ESPRESSO} (\texttt{QE}) code \cite{qe}. The effects of exchange and correlation are treated within the generalized gradient approximation proposed by Perdew, Burke, and Ernzerhof (GGA-PBE) \cite{pbe}. We use the ultrasoft pseudopotentials in \texttt{pslibrary.1.0.0} \cite{dalcorso}. The cutoff energies for the wavefunction and the charge density are 80 and 800 Ry, respectively. The self-consistent calculations within a spin-restricted approximation are performed by using a 20$\times$20$\times$1 $k$ grid and 20$\times$20$\times$20 $k$ grid for 2D and 3D structures, respectively \cite{MK}. The smearing parameter of Marzari-Vanderbilt \cite{smearing} is set to $\sigma=0.02$ Ry. For 2D structures, we set the size of the unit cell along the $c$ axis to be 14 \AA \ that is enough to avoid the interlayer coupling between different unit cells. The total energy and forces are converged within $10^{-4}$ Ry and $10^{-3}$ a.u., respectively.

We define the formation energy of a compound $XY$ in the structure $j$ as
\begin{eqnarray}
 E_j(XY) = \varepsilon_j(XY)- \frac{1}{2}\sum_{Q=X,Y} \min_j \varepsilon_j(QQ),
\end{eqnarray}
where the value of $\varepsilon_j(XY)$ is the total energy of $XY$ in the structure $j$ that is either BHC, BSQ, B2, L1$_0$, or B$_h$ in the present work, and the lowest energies of $\varepsilon_j(XX)$ and $\varepsilon_j(YY)$ among $j$ are subtracted. A negative value of $E_j(XY)$ indicates that forming a compound is energetically preferred. It has been known that the use of GGA-PBE underestimates $E_j$ for weakly bonded systems, whereas it gives accurate $E_j$ for strongly bonded systems \cite{nonlocalpbe,isaacs,ruzsinszky2019,ruzsinszky2020}. It has also been reported that the spin-orbit and van der Waals interactions do not play important roles in describing the stability of binary compounds \cite{ruzsinszky2019}.

We first optimize the lattice constant $a$ of $XY$ in the B2 structure. For the geometry optimization of the other structures, the initial guess for $a$ is set to be the value of $a$ optimized for the B2 structure. For BHC and BSQ, the buckling height $\delta$ is assumed to be $0.3a$ (total thickness $2\delta$). For computational efficiency, we will not study which of the high- and low-buckled structures are more stable, unless noted otherwise. For the L1$_0$ structure, the initial guess for $c/a$ is set to be 1.1 and 0.9. The lower energy structure is assigned to be L1$_0$ below. For the B$_h$ structure, the initial guess of $c/a$ is set to be the ideal value of $1.63$. Note that for some 2D compounds, no buckling structure ($\delta=0$) was found; and for some compounds, L1$_0$ with $c/a=1$ (i.e., B2) was obtained. During the optimization for BaPt in the B$_h$ structure, the hexagonal symmetry of the Bravais lattice was broken. For CuK and CuRb in the BSQ structure, CsCu in the L1$_0$ structure, and CsCu and CuRb in the B$_h$ structure, the self-consistent field calculations failed to converge. We also found that no Cu-Rb and Cs-Cu compounds were reposited in the Materials Project database \cite{materialsproject}. These may be related to the immiscibility of Cu into alkali metals in the bulk form. It would be desirable to understand the immiscibility in alloys more clearly using modern computational methods \cite{zhang_ML}.



The formation energies of 1081 compounds in the BHC, BSQ, B2, L1$_0$, and B$_h$ structures are provided in the SM \cite{SM}. In addition, the calculated $E_j$ for the 3D structures (B2, L1$_0$, and B$_h$) are compared with the reference values of 182 B2, 65 L1$_0$ and 96 B$_h$ compounds that are extracted from the Materials Project database \cite{materialsproject}. The agreement is, in general, good, except for binary compounds including magnetic elements such as Mn and Fe. 

The phonon dispersion calculations are performed within the density-functional perturbation theory \cite{dfpt} implemented in the \texttt{QE} code \cite{qe} and by using more than a 6$\times$6$\times$1 $q$ grid, that is, seven $q$ points for BHC and ten $q$ points for BSQ structures. When small imaginary frequencies are present around the $\Gamma$ point for a BHC structure, a denser grid is used (the cases of AgLi, GaLi, LiRh and MoRh), which can lead to positive values of frequencies around $\Gamma$. In the present study, the imaginary frequency of $\omega$ is represented by negative value.


For dynamically stable compounds in the BHC structure, we study the zero-point energy correction to $E_j$. When $E_{\rm BHC}$ is comparable to $E_{\rm BSQ}$, the vibrational free-energy contribution is also investigated \cite{grimvall2}. Our conclusions (i) and (ii) still hold when these effects are considered. Details are provided in the SM \cite{SM}.

\begin{figure}[tt]
\center
\includegraphics[scale=0.45]{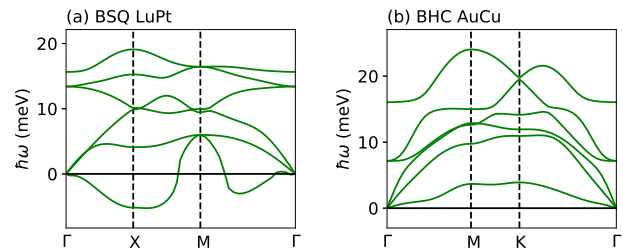}
\caption{The phonon dispersion curves of (a) BSQ LuPt and (b) BHC AuCu. } \label{fig2} 
\end{figure}


{\it Energetic stability and dynamical stability.} We identify only six BHC (AlPt, AuBa, AuLi, AuRb, LiPt, LuPt) and 13 BSQ (AlPt, AuBa, AuCa, AuCs, AuK, AuLi, AuRb, AuSr, BaPt, LuPt, PtSc, PtSr, PtY) structures having negative $E_j$, and find that these compounds except for BHC AlPt and LiPt are dynamically unstable. For example, BSQ LuPt has the lowest value of $E_{\rm BSQ}=-0.47$ eV among 19 compounds. Irrespective of this fact, the imaginary frequencies are observed around the point X and the middle of the $\Gamma$-M line, which confirms that instability, as shown in Fig.~\ref{fig2}(a).


Next, we focus on BHC AuCu. Although $E_{\rm BHC}({\rm AuCu})$ has a positive value (0.84 eV), BHC AuCu is dynamically stable as shown in Fig.~\ref{fig2}(b). This supports the recent experiment \cite{zagler}, where BHC AuCu is synthesized on a graphene substrate. Note that the use of the GGA-PBE functional \cite{pbe} to AuCu is known to yield an underestimation of the formation energy of ordered phases (L1$_0$ and L1$_2$) by a factor of 2. The use of a nonlocal functional is necessary to predict the value of the formation energy correctly \cite{nonlocalpbe}. It will be more important to study whether the substrate can produce a negative formation energy in this system, while such an investigation is beyond the scope of this work.   

Through the phonon dispersion calculations on BSQ LuPt and BHC AuCu, we point out that negative $E_j$ is neither a sufficient nor necessary condition for obtaining dynamically stable compounds in the structure $j$. Similar stability properties have been reported in silicene and germanene \cite{cahangirov} and 2D simple metals \cite{ono2020}.

\begin{figure}[tt]
\center
\includegraphics[scale=0.45]{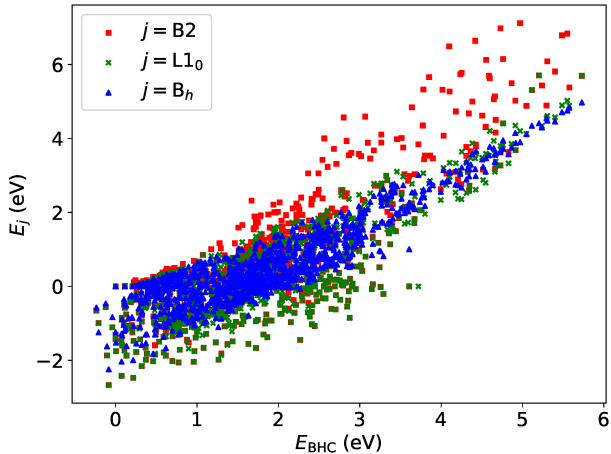}
\caption{Plots of $E_j(XY)$ for 3D structures $j$ as a function of $E_{\rm BHC}(XY)$.} \label{fig3} 
\end{figure}

\begin{table*}
\begin{center}
\caption{List of compounds that satisfy the conditions (C1) and (C2) described in text. ``SE'' and ``NR'' indicate that the B$_h$ structure has been ``synthesized experimentally'' and has ``not been reported'', respectively. ``DS'' and ``U'' indicate that the BHC structure is ``dynamically stable'' and ``unstable'', respectively. The figure in the parentheses is the number of compounds that match the conditions imposed. }
{
\begin{tabular}{ll}\hline\hline
 C1 / C2 \hspace{10mm} & Compounds \\
\hline
 B$_h$ SE / BHC DS \hspace{10mm} & IrLi \cite{varma_LiIr}, LiPd \cite{vucht_LiPd}, LiPt \cite{bronger_LiPt}, and LiRh \cite{sidhu_LiRh} (4)    \\ 
 B$_h$ SE / BHC U \hspace{10mm} & (0)     \\ 
 B$_h$ NR / BHC DS \hspace{10mm} & AgAl, AgAu, AgPd, AgPt, AuCd, CdMg, CrIr, CrRh, FeGa, GaMg, IrMo, IrRe, IrRu, IrTc, IrW \\
 & \hspace{0mm} IrZn, MgSn, MoRh, OsRe, OsRu, ReTc, RhTc, RuTc, and ScZr (24)    \\ 
 B$_h$ NR / BHC U \hspace{10mm} & AgZn, AlTi, AuTi, CaZn, GaTc, GaV, LiMg, MgZn, MoPt, OsV, PtRe, and PtTc (12)    \\ 
\hline\hline
\end{tabular}
}
\label{table1}
\end{center}
\end{table*}





\begin{figure}[tt]
\center
\includegraphics[scale=0.45]{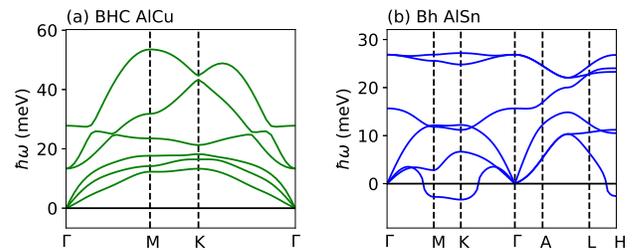}
\caption{The phonon dispersion curves of (a) BHC AlCu and (b) B$_h$ AlSn. } \label{fig4} 
\end{figure}

{\it Synthesizability and dynamical stability.} Although the energetic stability will not be a good indicator of dynamical stability, it can be used to study a similarity of compounds in between different structures \cite{nevalaita}. Figure \ref{fig3} shows the relationship of $E_j$ between BHC, B2, L1$_0$, and B$_h$ structures. The value of $E_{\rm BHC}$ is correlated to that of 3D structures. In particular, a strong correlation between the BHC and B$_h$ structures is observed: The correlation coefficients for the linear fit are 0.58 (B2), 0.59 (L1$_0$), and 0.78 (B$_h$). In the previous study \cite{ono2020}, we have demonstrated that AlCu in the B$_h$ structure is dynamically stable, while such a structure has not yet been synthesized. In the present study, we calculate the phonon dispersion curves of AlCu in the BHC structure. As shown in Fig.~\ref{fig4}(a), no imaginary frequencies are observed. This fact that both structures are dynamically stable must be due to the structural similarity between BHC and B$_h$, where both structures are constructed by stacking the hexagonal lattice. 

In order to study the metastability of B$_h$ compounds in the BHC structure, we choose binary compounds that can have a B$_h$ structure (space group of P$\bar{6}$m2), a negative formation energy, and zero band gap, referring to the Materials Project database \cite{materialsproject} and using the \texttt{pymatgen} code \cite{pymatgen}. We find that 40 compounds satisfy these conditions and identify that among them 28 compounds are dynamically stable. Table \ref{table1} lists the number of compounds satisfying the conditions: (C1) The B$_h$ structure has been synthesized experimentally and (C2) the BHC structure is dynamically stable. The Li-based compounds of IrLi, LiPd, LiPt, and LiRh satisfy both conditions (C1) \cite{varma_LiIr,vucht_LiPd,bronger_LiPt,sidhu_LiRh} and (C2). No compounds are found for satisfying only the condition (C1). Therefore, this establishes the structure-stability relationship: If a compound in the B$_h$ structure has been synthesized experimentally, that in the BHC structure is dynamically stable (B$_h\rightarrow$ BHC). We can also find that 12 compounds are unstable (see Table \ref{table1}), and those in the B$_h$ structure have not been synthesized experimentally \cite{materialsproject}. Therefore, the contraposition of ``B$_h\rightarrow$ BHC'' also holds. The phonon dispersions of these compounds in the BHC structure are provided in the SM \cite{SM}. 


We can confirm that the negatively large (or positively small) formation energy is neither a sufficient nor necessary condition for the dynamical stability of compounds again. The values of $E_{\rm BHC}$ for the stable Li-based compounds are as follows: For IrLi, LiPd, LiPt, and LiRh, $E_{\rm BHC}=0.67, 0.08, -0.21$, and $0.74$ eV, respectively. Among the 40 compounds listed in Table \ref{table1}, BHC FeGa has the largest value of $E_{\rm BHC}=2.06$ eV, irrespective of its dynamical stability. 



We investigate the dynamical stability of 46 Li-based compounds (Li$X$) in the BHC structure by calculating the phonon dispersions. We confirm that 15 BHC Li$X$ is dynamically stable when $X=$ group 2 (Sc, Y, and Lu), group 9 (Co, Rh, and Ir), group 10 (Ni, Pd, and Pt), group 11 (Cu and Ag), and group 13 (Al, Ga, and Tl) metals (BHC Li is also stable \cite{ono2020}), as provided in the SM \cite{SM}. This strongly suggests that these dynamical stabilities are correlated with the group in the periodic table.


When the condition of negative formation energy is not imposed in the search of B$_h$-type compounds \cite{pymatgen}, we find 96 compounds that consist of only metallic elements. In addition to the Li-based compounds mentioned above, only B$_h$ AlSn has already been synthesized experimentally \cite{kane_AlSn}. However, B$_h$ AlSn is not a compound but a solid solution. This fact gives rise to the instability of AlSn in the B$_h$ structure, as shown in Fig.~\ref{fig4}(b). We also investigate the dynamical stability of AlSn by considering three different BHC structures: no buckled, low-buckled, and high-buckled structures. As expected, all structures are unstable. The phonon dispersion curves in these structures are shown in the SM \cite{SM}. When the PBE functional revised for solids (PBEsol) \cite{pbesol} is used, AlSn in the B$_h$ and BHC structures are also unstable.


The present investigation implies that there would be two scenarios for the dynamical stability of compounds in the BHC structure. One is explained as a 2D analog of the B$_h$ structure (that is, the hexagonal symmetry is preferred), as in the four Li-based compounds listed in Table \ref{table1}. The other might be more complex: Although the ground state structure is different from the B$_h$ structure, a few or more metastable structures that can be correlated with the BHC structure are hidden in the potential energy surface, which can yield a dynamically stable BHC structure. While this is still a phenomenological discussion, we believe that the latter scenario explains the stability of the other 24 compounds (see Table \ref{table1}), AuCu \cite{zagler}, and AlCu \cite{ono2020}.  

In order to investigate the possibility that the BHC phase is transformed into the BSQ phase, we also study the dynamical stability of the BSQ phase for AlPt, AlCu, 15 Li-based compounds above, and 24 compounds listed in Table \ref{table1}. When BSQ is dynamically stable, we investigate the contribution from the vibrational free-energy. We confirm that no phase transition between BHC and BSQ is observed with increasing the temperature (see the SM for details \cite{SM}).


Before closing, we discuss how to synthesize 2D compounds experimentally. An appropriate substrate may be needed to create 2D systems that have either ordered, disordered, or more complex structures \cite{overbury,nielsen,yuhara_Rh,dhaka,yuhara_BiSn,yuhara_Ru,yuhara_Ag,yuhara_Al,sad,zagler}. The strain stored \cite{tersoff} and the coordination number of atoms \cite{nielsen,ono2021} around the surface will be modulated by the presence of the substrate, yielding stable 2D systems. For example, Pb and Sn, where these atoms are immiscible with each other, can form ordered structures on Rh(111) and Ru(0001) \cite{yuhara_Rh,yuhara_Ru}, disordered structures on Ag(111) \cite{yuhara_Ag}, and no mixed phase on Al(111) surfaces \cite{yuhara_Al}.

{\it Conclusion and future prospects.} We have demonstrated that (i) a negative formation energy is neither a sufficient nor a necessary condition for yielding dynamically stable 2D compounds, as in LuPt and AuCu; and (ii) given the synthesizability of a compound in the B$_h$ structure, the BHC structure is dynamically stable. We have identified 41 different binary compounds as candidates having the BHC structure (see the SM \cite{SM}). 

The present strategy for finding 2D compounds, relating the different structures in different dimensions, can be extended to other 2D structures. For example, the instability of LuPt (see Fig.~\ref{fig2}(a)) is due to the different ground state structure in the 3D crystal: Lu and Pt form the L1$_2$ structure. We consider that as a counterpart of L1$_2$, other 2D structures must be present. In this respect, the origin of the dynamical stability of 24 compounds listed in Table \ref{table1} as well as L1$_0$ AuCu, shown in Fig.~\ref{fig2}(b) and reported in Ref.~\cite{zagler}, remains unclear. In the present investigation, we have focused on compounds that consist of only metallic elements. We expect that the stability relationship between B$_h$ and BHC structures can also be applied to other B$_h$ compounds that include other elements such as C, N, and S (230 compounds or alloys \cite{pymatgen}) and that more analyses, with the help of 2D materials database \cite{ashton,choudhary,haastrup,feng}, can lead to an establishment of another stability-syhthesizability relationship. 


\begin{acknowledgments}
The authors thank M. Aoki for fruitful discussions. The computation was carried out using the facilities of the Supercomputer Center, the Institute for Solid State Physics, the University of Tokyo, and using the supercomputer ``Flow'' at Information Technologcy Center, Nagoya University.
\end{acknowledgments}



\end{document}